\documentclass[12pt,here]{article}
\usepackage{epsfig}

\begin{document}

\vspace*{-1cm}
\hspace*{10cm}
\vbox{
\hbox{ADP-01-33/T465}
}

\vspace*{1cm}

\begin{center}

{\Large
{\bf Complete analysis of spin structure function $g_{1}$ of $^3$He}}

\vspace*{1cm}
            
F.~Bissey$^{a,b,}$\footnote{fbissey@in2p3.fr}, V.~Guzey$^{a,}$\footnote{vguzey@physics.adelaide.edu.au}, M.~Strikman$^{c,}$\footnote{strikman@phys.psu.edu}, A.~Thomas$^{a,}$\footnote{athomas@physics.adelaide.edu.au}
     
\vspace*{1cm}

$^{a}$Special Research Centre for the Subatomic Structure of Matter (CSSM),  and Department of Physics and Mathematical Physics,\\ Adelaide University, Adelaide 5005, Australia\\
$^{b}$ Laboratoire de Physique Corpusculaire, Universit{\'e} Blaise Pascal, CNRS/IN2P3, 24 avenue des Landais, 63177 Aubi{\`e}re Cedex, France\\
$^{c}$Department of Physics, The Pennsylvania
State University,\\
 University Park, PA 16802, USA

\end{center}

\date{\today}

\vspace*{1cm}

\begin{abstract}

We present a comprehensive analysis of the nuclear effects important in DIS on polarized $^3$He over a wide range of Bjorken $x$, $10^{-4} \leq x \leq 0.8$. Effects relevant for the extraction of the neutron spin structure function, $g_{1}^{n}$, from the $^3$He data are emphasized.  

\end{abstract}

\section{Introduction}
\label{sec:intro}

One of the fundamental challenges of particle physics is to understand  the spin structure of protons, neutrons and nuclei in terms of their quarks and gluons. The main experimental tool, which is hoped to help answer the question, is deep inelastic scattering (DIS) of polarized leptons on polarized targets.

The present work is concerned with the spin structure functions $g_{1}^{^3{\rm He}}$ of the $^3$He nucleus and $g_{1}^{n}$ of the neutron. Since free neutron targets are not available, polarized deuterium and $^3$He are used as sources of polarized neutrons. The SMC experiments at CERN \cite{SMC} and the E143 \cite{E143} and E155 \cite{E155} experiments at SLAC employed polarized deuterium.
Polarized $^3$He was used by
the HERMES collaboration at DESY \cite{HERMES} and the E154 experiment at SLAC \cite{E154}.

Properties of protons and neutrons
 embedded in nuclei are expected to be different from those in free space. In particular, the neutron spin structure function $g_{1}^{n}$ is not equal to the $^3$He spin structure function $g_{1}^{^3{\rm He}}$ because of a variety of nuclear effects. These effects include spin depolarization,  nuclear binding and Fermi motion of nucleons,
the off-shellness of the nucleons, presence of
non-nucleonic degrees of freedom,
and  nuclear shadowing and antishadowing. 
While each of the above mentioned effects were considered in detail in the literature, no attempt was made to present a coherent and complete picture of all of them  in the entire range of Bjorken $x$. The aim of this work is to combine all the known results for the $^3$He structure function $g_{1}^{^3{\rm He}}$ in the range $10^{-4} \le x \le 0.8$ and to assess the importance of the nuclear effects on the extraction of the neutron structure function, $g_{1}^{n}$, from the $^3$He data.

\section{Spin depolarization, nuclear binding and Fermi motion}
\label{sec:sdep}

The nuclear effects of spin depolarization, binding and Fermi motion 
are traditionally described within the framework of the convolution approach \cite{Jaffe86}.
In this approximation, nuclear structure functions are given by the convolution of, in general, the 
off-shell
 nucleon structure functions with the light-cone nucleon momentum distributions. 
As a starting point, we assume that the structure functions of the
struck nucleon are those of the free and on-mass-shell nucleon  and that non-nucleonic degrees of freedom, such as vector mesons and the $\Delta$ isobar, do not contribute. In the following section we shall relax these assumptions.
The spin-dependent momentum distributions 
are given by the probability to 
find a nucleon 
with a given 
light-cone 
momentum fraction of the nucleus and the helicity of the nucleon aligned along  the helicity of the nucleus minus the probability that the helicities of the nucleon and the nucleus are opposite. 
In general, there is no unique procedure to obtain the light-cone nucleon momentum distributions from the non-relativistic nuclear wave function.
In what follows, we adopt the frequently used convention that the light-cone nucleon momentum distribution can be obtained from the nucleon spectral function \cite{CSPS93, SS93, Bissey01}. 
  Thus,
 $g_{1}^{^3{\rm He}}$ can be  represented as the convolution of the neutron ($g_{1}^{n}$) and proton ($g_{1}^{p}$) spin structure functions with the spin-dependent nucleon light-cone momentum distributions $\Delta f_{N/^3{\rm He}}(y)$, 
where $y$ is the ratio of the struck nucleon to nucleus light-cone plus components of the momenta 
\begin{equation}
g_{1}^{^3{\rm He}}(x,Q^2)=\int_{x}^{3} \frac{dy}{y} \Delta f_{n/^3{\rm He}}(y)g_{1}^{n}(x/y,Q^2)+\int_{x}^{3} \frac{dy}{y} \Delta f_{p/^3{\rm He}}(y)g_{1}^{p}(x/y,Q^2) \ .
\label{conv1}
\end{equation}
The motion of the nucleons inside the nucleus (Fermi motion) and their binding  are parametrized through the  distributions $\Delta f_{N/^3{\rm He}}$, which,
within the above discussed 
convention (one variant of the impulse approximation),
 can be readily calculated using the ground-state wave functions of $^3$He. Detailed calculations \cite{CSPS93, SS93, Bissey01}
by various groups using different ground-state wave function of $^3$He came to a similar conclusion that $\Delta f_{N/^3{\rm He}}(y)$ are sharply peaked around $y \approx 1$ 
due to the small average separation energy per nucleon. 
Thus, Eq.~(\ref{conv1}) is often approximated by
\begin{equation}
g_{1}^{^3{\rm He}}(x,Q^2)=P_{n} g_{1}^{n}(x,Q^2)+ 2P_{p} g_{1}^{p}(x,Q^2) \ .
\label{conv2}
\end{equation}
Here $P_{n}$ ($P_{p}$) are the effective polarizations of the neutron (proton) inside polarized $^3$He, which are defined by
\begin{equation}
P_{n,p}=\int_{0}^{3} dy \Delta f_{n,p/^3{\rm He}}(y) \ .
\label{pnp}
\end{equation}

In the first approximation to the ground-state wave function of $^3$He, only the neutron is polarized, which corresponds to the $S$-wave type interaction between any pair of the nucleons of $^3$He. In this case, $P_n$=1 and $P_p$=0. 
Realistic approaches to the wave function of $^3$He  include also  higher partial waves, notably the $D$ and $S^{\prime}$ partial waves, that arise due to the tensor component of the nucleon-nucleon force. This leads to the depolarization of spin of the neutron and polarization of protons  in $^3$He. The average of calculations with several models of nucleon-nucleon interactions and three-nucleon forces can be summarized
as  $P_{n}=0.86 \pm 0.02$ and $P_{p}=-0.028 \pm 0.004$ \cite{FGPBC}. The calculations of \cite{Bissey01} give similar values: $P_{n}=0.879$ and $P_{p}=-0.021$ for the PEST potential with 5 channels. We shall use  these values for $P_{n}$ and $P_{p}$ throughout this paper. 
One should note that most of the uncertainty in the values for $P_{n}$ and $P_{p}$ comes from the uncertainty in the $D$-wave of the $^3$He wave function. Thus, for the observables that are especially sensitive to the poorly constrained $P_{p}$, any theoretical predictions bear an uncertainty of at least 10\%. An example of such an observable is the point where the neutron structure function $g_{1}^{n}$ has a zero.

Equation~(\ref{conv1}) explicitly assumes that the nuclear spin structure function is given by the  convolution with the
 on-shell nucleon structure functions. In general, the nucleons bound together in a nucleus 
are subject to off-shell modifications
 so that the spin structure function of $^3$He, $g_{1}^{^3{\rm He}}$, should be expressed in terms of the off-shell nucleon spin  structure functions $\tilde{g}_{1}^{N}$ 
\begin{equation}
g_{1}^{^3{\rm He}}(x,Q^2)=\int_{x}^{3} \frac{dy}{y} \Delta f_{n/^3{\rm He}}(y)\tilde{g}_{1}^{n}(x/y,Q^2)+\int_{x}^{3} \frac{dy}{y} \Delta f_{p/^3{\rm He}}(y)\tilde{g}_{1}^{p}(x/y,Q^2) \ .
\label{conv3}
\end{equation}
In general, both $\Delta f_{N/^3{\rm He}}$ and $\tilde{g}_{1}^{N}$ in Eq.~(\ref{conv3}) depend on the virtuality of the struck nucleon.
However, in the region, where the Fermi motion effect is a small correction ($x \le 0.7$), one can substitute the off-shell nucleon structure functions by their values at some average virtuality. This was implicitly assumed in Eq.~(\ref{conv3}).

Off-shell corrections for such a light nucleus as $^3$He are not expected to be large. In this work, we use the results for $\tilde{g}_{1}^{n}$ and $\tilde{g}_{1}^{p}$ of \cite{Steffens}, where the off-shell corrections to valence parton distributions were estimated using the Quark Meson Coupling model \cite{QMC}. 
 The inclusion of the valence distributions only sets the lower limit of Bjorken $x$, where the result of \cite{Steffens} are applicable, to $x=0.2$.
 Also, 
since the Quark Meson Coupling model is based on the MIT bag model, 
the range of its validity is bound from above by $x \approx 0.7$.
Thus, we apply the results of \cite{Steffens} at $0.2 \leq x \leq 0.7$ and $Q^2 \leq 10$ GeV$^2$.

\begin{figure}[t]
\begin{center}
\epsfig{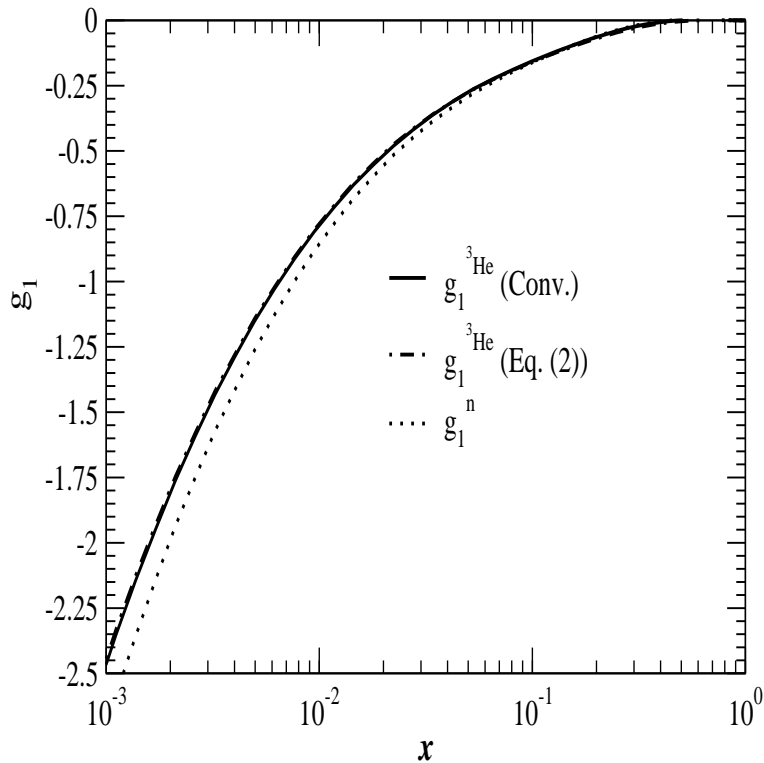}
\caption{The spin structure function $g_{1}^{^3{\rm He}}$ obtained with Eq.~(\ref{conv3}) (solid line) and Eq.~(\ref{conv2}) (dash-dotted line). The neutron structure function $g_{1}^{n}$ is shown as a dotted line.
} 
\label{fig:born}
\end{center}
\end{figure}

The results for the spin structure function $g_{1}^{^3{\rm He}}$ at $Q^2=4$ GeV$^2$ are presented in Fig.~\ref{fig:born}.
The solid curve depicts $g_{1}^{^3{\rm He}}$ obtained from Eq.~(\ref{conv3})
with $\Delta f_{N/^3{\rm He}}$ obtained using the PEST potential with 5 channels.
 This calculation includes all the nuclear effects
discussed so far: spin depolarization, Fermi motion and binding, and off-shell effects. We note that on a chosen logarithmic scale along the $x$ axis, the results of Eqs.~(\ref{conv3}) and (\ref{conv1}) are indistinguishable and shown by the solid curve. 
This  should be compared to the dash-dotted curve obtained from Eq.~(\ref{conv2}), which includes the spin depolarization effects only.
  Also, for comparison, the neutron spin structure function $g_{1}^{n}$ is given by the dotted line. 
The proton and neutron spin structure functions used in our calculations  were obtained using the standard, 
leading order, polarized parton distributions of Ref.~\cite{GRSV}.

We would like to stress that the small-$x$ nuclear effects ($10^{-4} \leq x \leq 0.2$), shadowing and antishadowing, were not taken into account so far. While we choose to present  our results in Fig.~\ref{fig:born} in the region $10^{-3} \leq x \leq 1$ and  to discuss our results in the region $10^{-4} \leq x \leq 0.8$ (see below), 
the most comprehensive expression for the $^3$He spin structure function, $g_{1}^{^3{\rm He}}$, is discussed in Sect.~\ref{sec:sha}.

As one can see from Fig.~\ref{fig:born},
the nuclear effects  discussed above, among which the most prominent one is 
nucleon spin depolarization, lead to a sizable difference between $g_{1}^{^3{\rm He}}$ and $g_{1}^{n}$. One finds that $g_{1}^{^3{\rm He}}$ is increased relative to $g_{1}^{n}$ by about 10\% for $10^{-4} \leq x \leq 0.01$. At larger $x$,
$0.01 < x \leq 0.25$, $g_{1}^{^3{\rm He}}$ and $g_{1}^{n}$ are equal with a few per cent accuracy. At $x > 0.3$
 both  
$g_{1}^{^3{\rm He}}$ and $g_{1}^{n}$ are very small so that while a quantitative comparison is possible, it is very sensitive to the details of the calculation. However,  
one can still make a weakly model-dependent statement that
at $x \approx 0.45$, where $g_{1}^{n}$ is extremely small because it changes sign, the contribution of $g_{1}^{p}$ to $g_{1}^{^3{\rm He}}$  becomes at least as important as that of $g_{1}^{n}$.

Also, it is important to assess how well Eq.~(\ref{conv2}) approximates the complete result of Eq.~(\ref{conv3}). In the region, where $x$ is 
small, 
$10^{-4} \leq x \leq 0.1$, Eq.~(\ref{conv2}) underestimates Eq.~(\ref{conv3}) by less than 1\%. However, for $x > 0.2$ the effect of convolution in Eq.~(\ref{conv3}) makes $g_{1}^{^3{\rm He}}$ sizably 
larger
 than predicted by Eq.~(\ref{conv2})
(see Fig.~\ref{fig:delta} emphasizing the large-$x$ region). 
Thus, ignoring for a moment the nuclear shadowing and antishadowing effects, Eq.~(\ref{conv2}) gives a very good approximation for   $g_{1}^{^3{\rm He}}$ over the range $10^{-4} \leq x \leq 0.1$. At larger $x$, the complete expression given in  Eq.~(\ref{conv3}) must be used.

Our conclusion that $g_{1}^{^3{\rm He}}$ can be approximated well by Eq.~(\ref{conv2}) only in the region $10^{-4} \leq x \leq 0.1$ is more stringent than the earlier result of \cite{CSPS93}, where the range of the applicability of Eq.~(\ref{conv2}) is $10^{-4} \leq x \leq 0.9$. As an argument in favor of the smaller range of the applicability of Eq.~(\ref{conv2}), we can consider the so-called EMC ratio for the unpolarized DIS on $^3$He. The deviation of the EMC ratio from unity is, like the deviation of the prediction of Eq.~(\ref{conv2}) from $g_{1}^{^3{\rm He}}$ based on Eq.~(\ref{conv1}), a measure of the Fermi motion and binding effects. It was shown in \cite{ABGKMPT} that the EMC ratio starts to deviate sizably from unity at $x >0.8$. In the work of \cite{SSS} this happens already for $x > 0.7$.

The convolution approach that forms the basis of Eqs.~(\ref{conv1},\ref{conv2},\ref{conv3})
 implies that the nuclear structure function can be obtained through convolution with   free and on-shell or off-shell nucleon structure functions. 
Using a reasonable model for the virtual photon-off-shell nucleon interaction,
 it was shown in \cite{PMT96} that the convolution approximation 
itself breaks down in  the region of relativistic kinematics, $x \geq 0.8$. Thus, $x=0.8$ defines the upper limit for the region of Bjorken $x$ studied in the present work.

It is customary to use Eq.~(\ref{conv2}) for the extraction of $g_{1}^{n}$ from $g_{1}^{^3{\rm He}}$ \cite{E155,HERMES,E154}. However, there are other nuclear effects that were not included in Eq.~(\ref{conv2}) that have also been shown to play an important role in polarized DIS on $^3$He. These effects include the presence of non-nucleon degrees of freedom and nuclear shadowing and antishadowing.

\section{Non-nucleonic degrees of freedom}
\label{sec:delta}

The description of the nucleus as 
 a mere
 collection of protons and neutrons is incomplete. In polarized DIS on the tri-nucleon system, this observation  can be  illustrated  by the following example \cite{FGS96}.
The Bjorken sum rule \cite{Bjorken} relates the difference of the  first moments of the proton and neutron spin structure functions to the axial vector coupling constant of  the neutron  $\beta$ decay $g_{A}$, where  $g_{A}=1.2670 \pm 0.0035$ \cite{Caso},
\begin{equation}
\int_{0}^{1}\Big(g_{1}^{p}(x,Q^2)-g_{1}^{n}(x,Q^2)\Big)dx =\frac{1}{6}g_{A}\Big(1+O(\frac{\alpha_{s}}{\pi})\Big) \ .
\label{delta1}
\end{equation}
Here the QCD radiative corrections are denoted as ``$O(\alpha_{s} / \pi)$''. This sum rule can be straightforwardly generalized 
to the $^3$He-$^3$H system:
\begin{equation}
\int_{0}^{3}\Big(g_{1}^{^3{\rm H}}(x,Q^2)-g_{1}^{^3{\rm He}}(x,Q^2)\Big)dx =\frac{1}{6}g_{A}|_{triton}\Big(1+O(\frac{\alpha_{s}}{\pi})\Big) \ ,
\label{delta2}
\end{equation}  
where $g_{A}|_{triton}$ is  the axial vector coupling constant of the triton $\beta$ decay, $g_{A}|_{triton}=1.211 \pm 0.002$ \cite{Budick}.
Taking the ratio of Eqs.\ (\ref{delta2}) and (\ref{delta1}), one obtains
\begin{equation}
\frac{\int_{0}^{3}\Big(g_{1}^{^3{\rm H}}(x,Q^2)-g_{1}^{^3{\rm He}}(x,Q^2)\Big)dx}{\int_{0}^{1}\Big(g_{1}^{p}(x,Q^2)-g_{1}^{n}(x,Q^2)\Big)dx}=\frac{g_{A}|_{triton}}{g_{A}}=0.956 \pm 0.004 \ .
\label{delta3}
\end{equation}
Note that the QCD radiative corrections cancel exactly in Eq.\ (\ref{delta3}).

Assuming  charge symmetry between the $^3$He and $^3$H ground-state wave functions, one can write the triton spin structure function $g_{1}(x,Q^2)$ in the form (see Eq.~({\ref{conv3}))
\begin{equation}
g_{1}^{^3{\rm H}}(x,Q^2)=\int_{x}^{3} \frac{dy}{y} \Delta f_{n/^3{\rm He}}(y)\tilde{g}_{1}^{p}(x/y,Q^2)+\int_{x}^{3} \frac{dy}{y} \Delta f_{p/^3{\rm He}}(y)\tilde{g}_{1}^{n}(x/y,Q^2) \ .
\label{delta4}
\end{equation}
Combining Eqs.~(\ref{conv3}) and (\ref{delta4})
and using the fact that, for example, 
\begin{eqnarray}
\int^{3}_{0} dx \int_{x}^{3} \frac{dy}{y} \Delta f_{n/^3{\rm He}}(y)\tilde{g}_{1}^{n}(x/y,Q^2)&=& \int^{3}_{0} dy \Delta f_{n/^3{\rm He}}(y) \int^{1}_{0} dx \tilde{g}_{1}^{n}(x,Q^2) \nonumber\\
&=&P_{n} \int^{1}_{0} dx \tilde{g}_{1}^{n}(x,Q^2) \ ,
\end{eqnarray} 
one obtains the following estimate for the ratio of the nuclear to nucleon Bjorken sum rules
\begin{equation}
\frac{\int_{0}^{3}\Big(g_{1}^{^3{\rm H}}(x,Q^2)-g_{1}^{^3{\rm He}}(x,Q^2)\Big)dx}{\int_{0}^{1}\Big(g_{1}^{p}(x,Q^2)-g_{1}^{n}(x,Q^2)\Big)dx}=\Big(P_n-2P_p\Big) \frac{\tilde{\Gamma}_{p}-\tilde{\Gamma}_{n}}{\Gamma_{p}-\Gamma_{n}}=0.921 \frac{\tilde{\Gamma}_{p}-\tilde{\Gamma}_{n}}{\Gamma_{p}-\Gamma_{n}}   \ .
\label{delta5}
\end{equation}
Here we used $P_{n}=0.879$ and $P_{p}=-0.021$;  
$\tilde{\Gamma}_{N}=\int^{1}_{0} dx \tilde{g}_{1}^{N}(x)$ and $\Gamma_{N}=\int^{1}_{0} dx g_{1}^{N}(x)$. 

If anything, the off-shell corrections of Ref.~\cite{Steffens} decrease
rather than increase  the bound nucleon spin structure functions (i.e. $(\tilde{\Gamma}_{p}-\tilde{\Gamma}_{n})/(\Gamma_{p}-\Gamma_{n}) < 1$). Thus,  
one can immediately see that the theoretical prediction for the ratio of the Bjorken sum rule for the $A=3$ and $A=1$ systems (Eq.~(\ref{delta5})), based solely on nucleonic degrees of freedom, underestimates the experimental result for the same ratio (Eq.~(\ref{delta3})) by 
 about 3.5\%.  This  demonstrates the need for new nuclear effects that are not included in Eqs.~(\ref{conv1},\ref{conv2},\ref{conv3}).

It has been known for a long time that non-nucleonic degrees of freedom, such as pions, vector mesons, the $\Delta$(1232) isobar, play an important role in the calculation of low-energy observables of nuclear physics. In particular, the analyses of Ref.~\cite{Saito} demonstrated that 
the two-body exchange currents involving a $\Delta$(1232) isobar increase the theoretical prediction for the axial vector coupling constant of triton by about 4\%,
which makes it consistent with experiment. Consequently, 
exactly the same mechanism must be present in case of deep inelastic scattering on polarized $^3$He and $^3$H. Indeed, as explained in Refs.~\cite{FGS96,BGST},
the direct correspondence between the calculations of the Gamow-Teller matrix element in the triton $\beta$ decay and the Feynman diagrams of DIS on $^3$He and $^3$H (see Fig.~1 of \cite{BGST}) requires that two-body exchange currents should play an equal role in both  processes. As a result, the presence of the $\Delta$ in the $^3$He and $^3$H wave functions should increase the ratio of Eq.~(\ref{delta5}) and make it consistent with Eq.~(\ref{delta3}).

The contribution of the $\Delta$(1232) to $g_{1}^{^3{\rm He}}$ is realized through Feynman diagrams involving the non-diagonal interference transitions $n \rightarrow \Delta^{0}$ and 
$p \rightarrow \Delta^{+}$. This requires new spin structure 
functions $g_{1}^{n \rightarrow \Delta^{0}}$ and $g_{1}^{p \rightarrow \Delta^{+}}$, as well as the effective polarizations $P_{n \rightarrow \Delta^{0}}$ and 
$P_{p \rightarrow \Delta^{+}}$. 
Taking into account the interference transitions, the spin structure functions 
$g_{1}^{^3{\rm He}}$ and $g_{1}^{^3{\rm H}}$ can be written as
\begin{eqnarray}
g_{1}^{^3{\rm He}}&=&\int_{x}^{3} \frac{dy}{y} \Delta f_{n/^3{\rm He}}(y)\tilde{g}_{1}^{n}(x/y,Q^2)+\int_{x}^{3} \frac{dy}{y} \Delta f_{p/^3{\rm He}}(y)\tilde{g}_{1}^{p}(x/y,Q^2) \nonumber\\
&+&2P_{n \rightarrow \Delta^{0}}g_{1}^{n \rightarrow \Delta^{0}}+4P_{p \rightarrow \Delta^{+}}g_{1}^{p \rightarrow \Delta^{+}} \ , \nonumber\\
g_{1}^{^3{\rm H}}&=&\int_{x}^{3} \frac{dy}{y} \Delta f_{n/^3{\rm He}}(y)\tilde{g}_{1}^{p}(x/y,Q^2)+\int_{x}^{3} \frac{dy}{y} \Delta f_{p/^3{\rm He}}(y)\tilde{g}_{1}^{n}(x/y,Q^2) \nonumber\\
&-&2P_{n \rightarrow \Delta^{0}}g_{1}^{p \rightarrow \Delta^{+}}-4P_{p \rightarrow \Delta^{+}}g_{1}^{n \rightarrow \Delta^{0}} \ .
\label{delta6}
\end{eqnarray} 
The minus sign in front of the interference terms in the expression for  $g_{1}^{^3{\rm H}}$ originates from the sign convention $P_{n \rightarrow \Delta^{0}} \equiv P_{n \rightarrow \Delta^{0}/^3{\rm He}}=-P_{p \rightarrow \Delta^{+}/^3{\rm H}}$ and 
$P_{p \rightarrow \Delta^{+}} \equiv P_{p \rightarrow \Delta^{+}/^3{\rm He}}=-P_{n \rightarrow \Delta^{0}/^3{\rm H}}$.
 
The interference structure functions can be related to $g_{1}^{p}$ and $g_{1}^{n}$ within the quark parton model using the general structure of the SU(6) wave functions \cite{Boros}
\begin{equation}
g_{1}^{n \rightarrow \Delta^{0}}=g_{1}^{p \rightarrow \Delta^{+}}=\frac{2 \sqrt{2}}{5}\Big(g_{1}^{p}-4 g_{1}^{n} \Big) \ .
\label{su6}
\end{equation}
This simple relationship is valid in the range of $x$ and $Q^2$ where the contribution of sea quarks and gluons to $g_{1}^{N}$ can be safely omitted, i.e. at $0.5 \leq Q^2 \leq 5$ GeV$^2$ and $0.2 \leq x \leq 0.8$ if the parametrization of Ref.~\cite{GRSV} is used. 

In principle, the effective polarizations of the interference contributions $P_{n \rightarrow \Delta^{0}}$ and $P_{p \rightarrow \Delta^{+}}$ can be calculated using a $^3$He wave function that includes the $\Delta$ resonance. This is an involved computational problem. Instead, we chose to find $P_{n \rightarrow \Delta^{0}}$ and $P_{p \rightarrow \Delta^{+}}$ by {\it requiring} that the
use of the $^3$He and $^3$H structure functions of Eq.~(\ref{delta6}) 
 gives the experimental ratio of the nuclear to nucleon Bjorken  sum rules (\ref{delta3}). Substituting Eq.~(\ref{delta6}) into Eq.~(\ref{delta3}) yields 
\begin{equation}
-2\Big(P_{n \rightarrow \Delta^{0}}+2P_{p \rightarrow \Delta^{+}}\Big)\frac{\int_{0}^{1} dx \Big(g_{1}^{n \rightarrow \Delta^{0}}(x)+g_{1}^{p \rightarrow \Delta^{+}}(x)\Big)}{\Gamma_{p}-\Gamma_{n}}=0.956-0.921 \frac{\tilde{\Gamma}_{p}-\tilde{\Gamma}_{n}}{\Gamma_{p}-\Gamma_{n}}\ .
\label{mok}
\end{equation} 
Next, we use Eq.~(\ref{su6}) to relate the interference structure functions to the off-shell modified proton and neutron spin structure functions. The latter are proportional to the on-shell nucleon spin spin structure function in the model of Ref.~\cite{Steffens}. Thus, using the parametrization of \cite{GRSV} one can find the first moments $\tilde{\Gamma}_{N}$ and $\Gamma_{N}$. 
At $Q^2=4$ GeV$^2$, we obtain $\Gamma_{p}=0.151$ and $\Gamma_{n}=-0.060$ for on-shell nucleons; $\tilde{\Gamma}_{p}=0.147$ and $\tilde{\Gamma}_{n}=-0.060$ for off-shell nucleons (when the off-shell effects are present in the range $0.2 \leq x \leq 0.7$).

Using Eqs.~(\ref{su6},\ref{mok}) and the calculated first moments,
we find for the necessary combination of the effective polarizations:
\begin{equation}
2\Big(P_{n \rightarrow \Delta^{0}}+2P_{p \rightarrow \Delta^{+}}\Big)=-0.025 \ .
\label{effpol}
\end{equation}
Note that Eq.~(\ref{effpol}) gives a value that is very similar to the one reported in our original publication \cite{BGST}.

Equations~(\ref{delta6},\ref{su6},\ref{effpol}) enable one to write an explicit expression for the $^3$He spin structure function, which takes into account the additional Feynman diagrams corresponding to the non-diagonal interference $n \rightarrow \Delta^{0}$ and  $p \rightarrow \Delta^{+}$ transitions (see Fig.~1 of \cite{BGST}) and which complies with the experimental value of the ratio of the Bjorken sum rules~(\ref{delta3}): 
\begin{eqnarray}
g_{1}^{^3{\rm He}}&=&\int_{x}^{A} \frac{dy}{y} \Delta f_{n/^3{\rm He}}(y)\tilde{g}_{1}^{n}(x/y,Q^2)+\int_{x}^{A} \frac{dy}{y} \Delta f_{p/^3{\rm He}}(y)\tilde{g}_{1}^{p}(x/y,Q^2) \nonumber\\
&-&0.014\Big(\tilde{g}_{1}^{p}(x,Q^2)-4\tilde{g}_{1}^{n}(x,Q^2)\Big)  \ .
\label{delta7}
\end{eqnarray}
Note that
in our model for the contribution of the $\Delta$ isobar to $g_{1}^{^3{\rm He}}$,
  the last term in Eq.~(\ref{delta7}) 
is strictly valid and included
only in the region $0.2 \leq x \leq 0.8$.

The results of the calculation of $g_{1}^{^3{\rm He}}$ at $Q^2=4$ GeV$^2$ based on 
 Eq.~(\ref{delta7}) are presented in Fig.~\ref{fig:delta} as a solid curve. 
They should be compared to $g_{1}^{^3{\rm He}}$ obtained from Eq.~(\ref{conv3}) (dash-dotted curve) and to  $g_{1}^{^3{\rm He}}$ obtained from Eq.~(\ref{conv2}) (dashed line). The neutron spin structure function,
$g_{1}^{n}$, is given by the dotted curve.
\begin{figure}[t]
\begin{center}
\epsfig{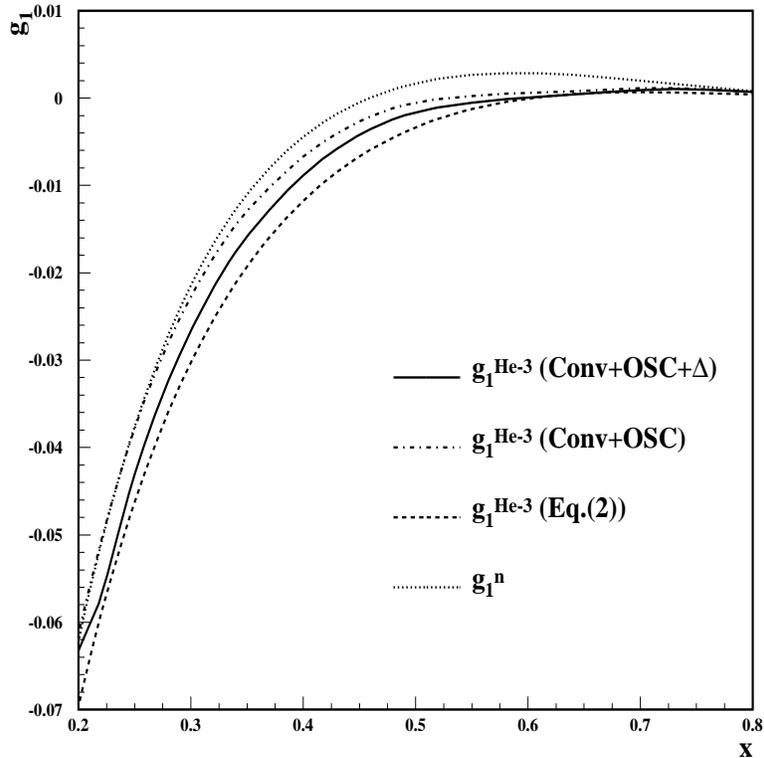}
\caption{The spin structure function $g_{1}^{^3{\rm He}}$ obtained from Eq.~(\ref{delta7}) (solid curve), Eq.~(\ref{conv3}) (dash-dotted curve), and Eq.~(\ref{conv2}) (dashed curve). The free neutron spin structure function $g_{1}^{n}$ is shown by the dotted curve. For all curves $Q^2$=4 GeV$^2$.}
\label{fig:delta}
\end{center}
\end{figure}
One can see from Fig.~\ref{fig:delta} that the presence of the $\Delta$(1232) isobar in the $^3$He wave function works to decrease $g_{1}^{^3{\rm He}}$ relative to the prediction of Eq.~(\ref{conv3}). This decrease is 12\% at $x=0.2$ and increases at larger $x$, peaking for $x \approx 0.46$, where $g_{1}^{n}$ changes  sign.

Equation~(\ref{delta7}) describes the nuclear effects of the nucleon spin depolarization and the presence of non-nucleon degrees of freedom in the $^3$He ground-state wave function and is based on the convolution formula~(\ref{conv1}). Since the convolution formalism implies {\it incoherent} scattering off 
nucleons 
and nucleon resonances
 of the target, coherent nuclear effects present at small values of Bjorken $x$ are ignored. In the next section we demonstrate the role played by two coherent effects, nuclear shadowing and antishadowing, in DIS on polarized $^3$He.

\section{Nuclear shadowing and antishadowing}
\label{sec:sha}

At high energies or small Bjorken $x$, the virtual photon can interact coherently with several nucleons in the nuclear target. This is manifested in a specific behaviour of nuclear structure functions that cannot be accommodated by the 
convolution approximation. In particular, by studying DIS of muons on a range of unpolarized nuclear targets, the NMC collaboration \cite{NMC} demonstrated that 
the ratio $2F_{2}^{A}/(AF_{2}^{D})$ deviates significantly from unity: it is 
smaller than unity for $0.0035 \leq x \leq 0.03-0.07$ and  is larger than unity for $0.03-0.07 \leq x \leq 0.2$. The depletion of the ratio  $2F_{2}^{A}/(AF_{2}^{D})$ is called nuclear shadowing, while the enhancement is termed nuclear antishadowing. Both of the effects break down the convolution approximation. 

Quite often nuclear targets are  used in polarized DIS experiments. While these experiments  do not  reach such low values of $x$ as the unpolarized fixed target experiments, where nuclear shadowing is important, the antishadowing region is still covered. 
In the absence of a firm theoretical foundation, nuclear shadowing and antishadowing have been completely ignored in the analysis of the DIS data on polarized nuclei. The prime motivation of this section is to demonstrate that these two effects are quite significant and do affect the extraction of the nucleon spin  functions  from the nuclear data.

The physical picture of nuclear shadowing in DIS is especially transparent in the target rest frame. At high energy, the incident photon 
interacts with hadronic targets by fluctuating into hadronic configurations  
$|h_{k} \rangle$, long before it hits the target:
\begin{equation}
|\gamma^{\ast} \rangle =\sum_{k} \langle h_{k}|\gamma^{\ast}\rangle |h_{k} \rangle \ ,
\label{sh1}
\end{equation}
where ``{\it k}'' is a generic label for the momentum and helicity of the hadronic fluctuation $h_{k}$. 
Thus, the total cross section for virtual photon-nucleus scattering can be presented in the general form
\begin{equation}
\sigma^{tot}_{\gamma^{\ast} A}=\sum_{k} |\langle h_{k}|\gamma^{\ast} \rangle|^2 \sigma_{h_{k}A}^{tot} \ .
\label{sh2}
\end{equation}
Here $|\langle h_{k}|\gamma^{\ast} \rangle|^2$ is the probability of the fluctuation $|\gamma \rangle \to |h_{k} \rangle$. In obtaining Eq.~(\ref{sh2}) from Eq.~(\ref{sh1}) we assumed that the fluctuations $h_{k}$ do not mix during the interaction. In general, this is not true since  various configurations $|h_{k} \rangle$ contribute to the expansion~(\ref{sh1}) and those states are not eigenstates of the scattering matrix, i.e. they  mix. However, one can replace the  series~(\ref{sh1}) by an effective state $|h_{eff} \rangle$ with the  mass $M_{eff}^2 \approx Q^2$ that interacts with the nucleons of the nuclear target with the effective cross section $\sigma_{eff}$. Within such an  approximation, Eq.~(\ref{sh2}) is valid and becomes
\begin{equation}
\sigma^{tot}_{\gamma^{\ast} A}=|\langle h_{eff}|\gamma^{\ast} \rangle|^2 \sigma_{h_{eff}A}^{tot} \ .
\label{sh2b}
\end{equation}  
Since the effective hadronic fluctuation $h_{eff}$ can interact coherently with several nucleons of the target, $\sigma_{h_{eff}A}^{tot} < A \sigma_{h_{eff}N}^{tot}$, which leads to  $\sigma^{tot}_{\gamma^{\ast} A} < A \sigma^{tot}_{\gamma^{\ast} N}$ and to  shadowing of the nuclear structure functions.
The approximation by a single effective state (see Eq.~(\ref{sh2b})) was used to estimate  the nuclear shadowing correction to spin structure functions of  deuterium \cite{deuterium}, $^3$He \cite{FGS96, GS99}, $^7$Li \cite{GS99}, and $^6$LiD \cite{li6d}.

By definition, the spin structure function $g_{1}^{^3{\rm He}}$ can be expressed as 
\begin{equation}
g_{1}^{^3{\rm He}} \propto \sigma^{\uparrow \downarrow}_{\gamma^{\ast} A}-\sigma^{\uparrow \downarrow}_{\gamma^{\ast} A} \propto
 \sigma_{h_{eff}A}^{\uparrow \uparrow} - \sigma_{h_{eff}A}^{\uparrow \downarrow} \ ,
\label{g1}
\end{equation}
where $\sigma_{h_{eff}A}^{\uparrow \uparrow}$ ($\sigma_{h_{eff}A}^{\uparrow \downarrow}$) is the cross section for the scattering when the helicities of the projectile  and the nucleus are parallel (antiparallel).
The cross sections $\sigma_{h_{eff}A}^{\uparrow \uparrow}$ and $\sigma_{h_{eff}A}^{\uparrow \downarrow}$ can be calculated using the standard Gribov-Glauber multiple scattering formalism. Within this approach, $\sigma_{h_{eff}A}^{\uparrow \uparrow}$ and $\sigma_{h_{eff}A}^{\uparrow \downarrow}$ receive contributions from the virtual photon scattering on each nucleon, each pair of nucleons and all three nucleons of the target. The first kind of contribution corresponds to incoherent scattering on the nucleons and leads to  $g_{1}^{^3{\rm He}}$ as given by Eq.~(\ref{conv3}). The simultaneous, coherent scattering on pairs of  nucleons and all three of them results in the shadowing correction to $g_{1}^{^3{\rm He}}$, $\delta g_{1}^{^3{\rm He}}$. Detailed calculations of $\delta g_{1}^{^3{\rm He}}$ are presented in Appendix~A.  
Thus, including the nuclear shadowing correction, the spin structure function of $^3$He reads
 \begin{eqnarray}
&&g_{1}^{^3{\rm He}}=\int_{x}^{3} \frac{dy}{y} \Delta f_{n/^3{\rm He}}(y)\tilde{g}_{1}^{n}(x/y)+\int_{x}^{3} \frac{dy}{y} \Delta f_{p/^3{\rm He}}(y)\tilde{g}_{1}^{p}(x/y) \nonumber\\
&&-0.014\Big(\tilde{g}_{1}^{p}(x)-4\tilde{g}_{1}^{n}(x)\Big)+a^{sh}(x) g_{1}^{n}(x)+b^{sh}(x) g_{1}^{p}(x)  \ ,
\label{g1sh}
\end{eqnarray}   
where $a^{sh}$ and $b^{sh}$ are functions of $x$ and $Q^2$ and are calculated using a particular model for $\sigma_{eff}$ and a specific form of the $^3$He ground-state wave function.

The present accuracy of fixed target polarized DIS experiments on nuclear targets is not sufficient for dedicated studies of nuclear shadowing. Thus, one can only use information obtained from  unpolarized DIS on nuclei. All of those experiments -- NMC at CERN, E139 at SLAC, BCDMS and E665 at Fermilab -- demonstrated that nuclear shadowing at $10^{-4} \leq x \leq 0.03-0.07$ is followed by some antishadowing at $0.03-0.07 \leq x \leq 0.2$. It is natural to assume a similar pattern
 for polarized DIS on $^3$He. Thus, Eq.~(\ref{g1sh}) can  be generalized as
\begin{eqnarray}
&&g_{1}^{^3{\rm He}}=\int_{x}^{3} \frac{dy}{y} \Delta f_{n/^3{\rm He}}(y)\tilde{g}_{1}^{n}(x/y)+\int_{x}^{3} \frac{dy}{y} \Delta f_{p/^3{\rm He}}(y)\tilde{g}_{1}^{p}(x/y) \nonumber\\
&&-0.014\Big(\tilde{g}_{1}^{p}(x)-4\tilde{g}_{1}^{n}(x)\Big)+a(x) g_{1}^{n}(x)+b(x) g_{1}^{p}(x)  \ ,
\label{g1sh2}
\end{eqnarray}
where $a$ ($b$) coincide with $a^{sh}$ ($b^{sh}$) in the nuclear shadowing region of Bjorken $x$ and model antishadowing at larger $x$. Since the shadowing contribution in Eq.~(\ref{g1sh}) breaks the equivalence of the theoretical and experimental values for the ratio of the nuclear to nucleon Bjorken sum rules, one can reinstate the equivalence by a suitable choice of antishadowing. Thus, we model antishadowing by requiring that Eq.~(\ref{g1sh2}) and its $^3$H counterpart give the correct ratio in Eq.~(\ref{delta3}).  
Substituting Eq.~(\ref{g1sh2}) into Eq.~(\ref{delta3}), we obtain the following condition on  
the functions 
$a$  and $b$
\begin{equation}
\int_{10^{-4}}^{0.2} dx \Big(a(x)-b(x)\Big) \Big(g_{1}^{p}(x)-g_{1}^{n}(x)\Big)=0 \ .
\label{anti}
\end{equation}
Note that the the lower limit of integration, $x=10^{-4}$, is somewhat artificial since it is defined by the range of $x$ covered by the parametrizations of $g_{1}^{p}$ and   $g_{1}^{n}$ of Ref.~\cite{GRSV}.
In general, the functions $a$  and $b$ are independent.
In order to simplify the modelling of $a$  and $b$ in the antishadowing region, we assume that they are proportional each other, i.e.
 $a(x)=c b(x)$, where $c$ is a constant.
Our calculations of $a$  and $b$ in the nuclear shadowing region justify this assumption with high accuracy and enable us to fix the value for the constant $c$: $c=57$. 
The value of the coefficient $c$ reflects the dominance of the effective polarization of the neutron, $P_{n}$, over that  of the proton, $P_{p}$.

As explained in detail in the Appendix, in calculating the shadowing correction $\delta g_{1}^{^3{\rm He}}$ and $a^{sh}$ and $b^{sh}$ entering Eq.~(\ref{g1sh}) we used two 
versions of the model for $\sigma_{eff}$, which  was formulated in the work by Frankfurt and Strikman \cite{FS99}. 
 In this model,
$\sigma_{eff}$ is inferred  using a connection between the nuclear shadowing correction to the nuclear structure function $F_{2}^{A}$ and the proton diffractive structure function $F_{2}^{D}$. Both structure functions, $F_{2}^{A}$ and 
$F_{2}^{D}$, enter unpolarized DIS. However, we still choose to use the corresponding $\sigma_{eff}$ to evaluate nuclear shadowing in polarized DIS. In principle, if the data on polarized electron-proton diffraction existed, one could readily improve $\sigma_{eff}$ necessary for the calculation of nuclear shadowing in polarized DIS on nuclei, using the formalism developed in Ref.~\cite{FS99}.
One of the main reasons why we decided to use the results of Ref.~\cite{FS99} is because its $\sigma_{eff}$
 corresponds to the leading twist shadowing correction to the nuclear parton densities, i.e $\sigma_{eff}$ decreases logarithmically with $Q^2$ according to the QCD evolution equation. We are forced to use the leading twist model of nuclear shadowing because in order to model the antishadowing contribution, we will use the Bjorken sum rule, which is a leading twist result.

Alternatively, if we were not concerned with leading twist shadowing, we could use another model for $\sigma_{eff}$. For example, the data on inclusive nuclear structure functions were successfully described within the two-phase model of Refs.~\cite{MT}. This model contains both the leading twist (Pomeron and triple  Pomeron exchanges) and subleading twist (vector meson) contributions. The latter contribution is required to describe the data at low $x$ and low $Q^2$, where higher twist effects are expected to be important. Thus, in applying shadowing corrections to low-$Q^2$ data points (such as the HERMES data used in our analysis), one should be aware of the higher twist effects, which will make predictions less model-independent.

Nuclear shadowing is followed by some antishadowing. The cross-over point between the two regions, $x_{0}$, is 
a parameter of the model of Ref.~\cite{FS99} ($\sigma_{eff}$ becomes zero at $x_{0}$), which should be inferred from experiment. Unfortunately, even the most precise NMC data \cite{NMC} is inconclusive about the exact position of the cross-over point $x_{0}$: experimental errors allow $x_{0}$ to be positioned  anywhere between 0.03 and 0.07. In order to take into account this ambiguity, which constitutes major theoretical uncertainty of our treatment of antishadowing, we considered two extreme versions of $\sigma_{eff}$ vanishing at $x_{0}=0.03$ and $x_{0}=0.07$.

Results for the function $a$ calculated with the both versions of $\sigma_{eff}$ are presented in Fig.~\ref{fig:ab} at  $Q^2=4$ GeV$^2$. 
\begin{figure}[h]
\begin{center}
\epsfig{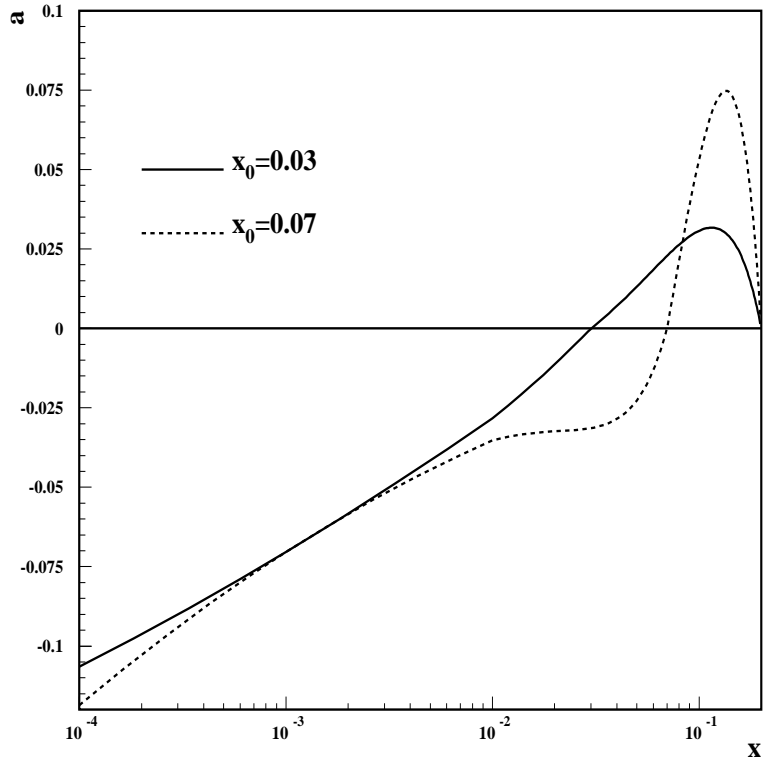}
\caption{The coefficient $a$ entering Eq.~(\ref{g1sh2}) that describes nuclear shadowing and antishadowing corrections. The solid curve corresponds to $x_{0}=0.03$; the dashed curve corresponds to $x_{0}=0.07$.}
\label{fig:ab}
\end{center}
\end{figure}
In both cases the amount of nuclear shadowing at small $x$ is 
quite similar: at $x=10^{-4}$, the shadowing correction amounts to 11\%, when $x_{0}=0.03$, and to 12\%, when $x_{0}=0.07$. These results are consistent with 
 the earlier results of Refs.~\cite{FGS96, GS99}, where the shadowing correction to $g_{1}^{^3{\rm He}}$ was of the order 10\%. Moreover, such a good consistency between the present calculation using the exact wave function of $^3$He  and the calculations using a simple Gaussian shape for the $^3$He wave function, where only the neutron was polarized ~\cite{FGS96, GS99}, demonstrates  that higher partial waves ($S^{\prime}$ and $D$) are unimportant in the calculation of the shadowing correction for polarized $^3$He. 

By choosing two different cross-over points consistent with the experimental data, we can assess the theoretical uncertainty of our modelling of antishadowing.
Since $a^{sh}$ in the model with  the cross-over point $x_{0}=0.07$ occupies a narrower region of $x$, the corresponding $a$ in the antishadowing  region reaches higher values relative to the model with the cross-over point $x_{0}=0.03$. For instance, at its maximum the antishadowing correction is of the order 3\%, when $x_{0}=0.03$, and of the order 7\%, when $x_{0}=0.07$. These values for the antishadowing correction are significantly smaller than those reported in \cite{FGS96, GS99}. This discrepancy must have arisen from 
slightly different shapes of the $x$-dependence of  antishadowing and different parametrizations for $g_{1}^{p}$ and $g_{1}^{n}$, which enter Eq.~(\ref{anti}) and determine the magnitude of antishadowing.   

One should note that our approach to antishadowing based on the ratio of the Bjorken sum rules (see Eq.~(\ref{delta3})) is just one of several ways to treat  antishadowing. In unpolarized DIS on nuclei, other models of antishadowing include the model of \cite{Brodsky}, where antishadowing explained by introducing both the Pomeron and  Reggeon exchanges (there is only the Pomeron exchange in the present work)
for the virtual photon-nucleon interaction, or the model of \cite{Melnitchouk}, where antishadowing is a consequence of the virtual photon scattering off the
pion cloud of the nucleus.

Using our calculations for the coefficients $a$ and $b$, we present the most comprehensive result for the $^3$He spin structure function $g_{1}^{^3{\rm He}}$ based on Eq.~(\ref{g1sh2}) in Fig.~\ref{fig:all}.
\begin{figure}[t]
\begin{center}
\epsfig{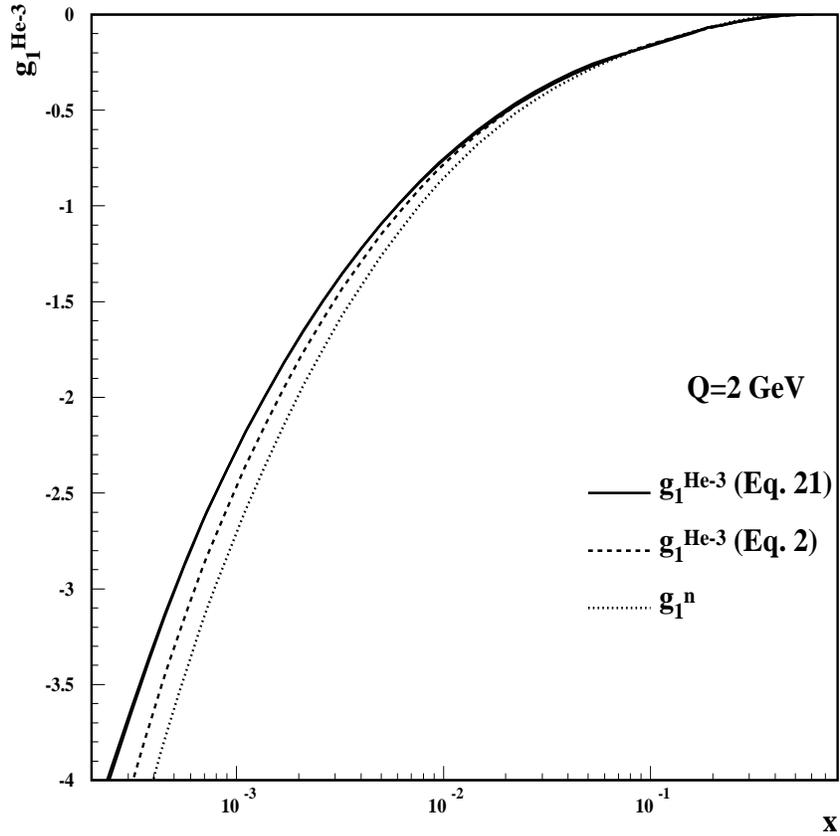}
\caption{The full calculation of $g_{1}^{^3{\rm He}}$ including nuclear shadowing and antishadowing based on Eq.~(\ref{g1sh2}) (two solid curves) compared to the result of Eq.~(\ref{conv2}) (dashed curve) and to $g_{1}^{n}$ (dotted curve).}
\label{fig:all}
\end{center}
\end{figure}
The solid curve includes all of the effects discussed above: nucleon spin depolarization,  Fermi motion and binding effects, the presence of the $\Delta$ isobar in the $^3$He wave function, and nuclear shadowing and antishadowing. On the chosen scale, the results of the calculations with the two different cross-over points $x_{0}$ are indistinguishable and are shown by the same solid curve. This
should be compared to the calculation of  $g_{1}^{^3{\rm He}}$ based on Eq.~(\ref{conv2}) (dashed curve) and to the free neutron spin structure function $g_{1}^{n}$ (dotted curve).

The comparison between the solid and the dashed curves is very important 
and constitutes one of the main results of the present work. So far, in the analysis of all experiments on polarized DIS on polarized $^3$He -- the E142 and E154 experiments at SLAC and the HERMES experiment at DESY -- it was assumed that the $^3$He spin structure function  $g_{1}^{^3{\rm He}}$ can be represented well by 
Eq.~(\ref{conv2}). However, the sizable difference between the full calculation based on Eq.~(\ref{g1sh2}) and the one based on Eq.~(\ref{conv2}) indicates that it is important to treat 
all the relevant nuclear effects
equally carefully. In the nuclear shadowing region, $10^{-4} \leq x \leq x_{0}$,  
$g_{1}^{^3{\rm He}}$ based on Eq.~(\ref{g1sh2}) is larger than that based on 
Eq.~(\ref{conv2}). 
For example, at $x=10^{-3}$ the difference is 11\% for the calculation with $\sigma_{eff}$  of Ref.~\cite{FS99} with $x_{0}=0.03$ 
 and 13\% for the the calculation with $x_{0}=0.07$. 
In the antishadowing region, $x_{0} < x \leq 0.2$, $g_{1}^{^3{\rm He}}$ based on Eq.~(\ref{g1sh2}) is smaller than the one predicted by Eq.~(\ref{conv2}). 
The difference can be read off from the corresponding curves for the function $a$ from Fig.~\ref{fig:ab}. For instance, for the calculation with $x_{0}=0.07$, the full result for $g_{1}^{^3{\rm He}}$ is smaller than the approximate one of Eq.~(\ref{conv2}) by 6\% at $x=0.13$.
Since nuclear shadowing and antishadowing are absent at $x >0.2$, Eq.~(\ref{g1sh2}) coincides with Eq.~(\ref{delta3}) in this region and for the comparison between the full calculations and an approximate one given by Eq.~(\ref{conv2}) we refer the reader to the discussion of Fig.~\ref{fig:delta}.

\section{Extraction of $g_{1}^{n}$ from the $^3$He data}

In the previous section we presented the calculation of the spin structure function of $^3$He, $g_{1}^{^3{\rm He}}$, which includes the effects of nuclear shadowing and antishadowing, the presence of the $\Delta$(1232) isobar in the $^3$He wave function, nucleon spin depolarization, Fermi motion and binding, and off-shellness of the nucleons. The resulting $g_{1}^{^3{\rm He}}$ given by Eq.~(\ref{g1sh2}) deviates from the approximate expression for  $g_{1}^{^3{\rm He}}$ given by Eq.~(\ref{conv2}), which takes into account only the effect of the nucleon spin depolarization. Since 
Eq.~(\ref{conv2}) was used to extract the neutron spin structure function $g_{1}^{n}$ from $g_{1}^{^3{\rm He}}$, one should reanalyse the data using the complete Eq.~(\ref{g1sh2}). In particular, we present our corrections to $g_{1}^{n}$ obtained from DIS on polarized $^3$He by the E154 Collaboration at SLAC \cite{E154} and the HERMES Collaboration at DESY \cite{HERMES}.

Let us denote the neutron structure function obtained from $g_{1}^{^3{\rm He}}$, using Eq.~(\ref{conv2}), as $g_{1 \, {\rm exp}}^{n}$. On the other hand, neglecting  Fermi motion, binding and off-shell effects, Eq.~(\ref{g1sh2}) relates  $g_{1}^{^3{\rm He}}$ to the true free neutron spin structure function $g_{1}^{n}$:
\begin{equation}
g_{1}^{^3{\rm He}}=P_{n}g_{1}^{n}+2P_{p}g_{1}^{p}-0.014\Big(g_{1}^{p}(x)-4g_{1}^{n}(x)\Big)+a(x) g_{1}^{n}(x)+b(x) g_{1}^{p}(x)  \ .
\label{true}
\end{equation}
Thus, the influence of the effects of nuclear shadowing and antishadowing and the $\Delta$ isobar on the $g_{1}^{n}$ extracted from the $^3$He data can be represented by the ratio of $g_{1}^{n}$ based on Eq.~(\ref{true}) to $g_{1 \, {\rm exp}}^{n}$
\begin{equation}
\frac{g_{1}^{n}}{g_{1 \, {\rm exp}}^{n}}=\frac{P_{n}+g_{1}^{p}/g_{1 \, {\rm exp}}^{n}\Big(0.014-b(x)\Big)}{P_{n}+0.056+a(x)} \ .
\label{ratio}
\end{equation}
Note that the coefficients $0.014$ and $0.056$ should be set to zero for $x < 0.2$ and $x>0.8$. By definition, the coefficients $a$ and $b$ are equal to  zero for $x > 0.2$.

The results of the application of Eq.~(\ref{ratio}) to $g_{1 \, {\rm exp}}^{n}$ reported by the E154 and HERMES Collaborations are presented in Fig.~\ref{fig:ratio}. 
We present calculations for the case, when $x_{0}=0.07$.
For simplicity we assumed that the functions $a$ and $b$ entering Eq.~(\ref{ratio}) and describing the amount of nuclear shadowing and antishadowing do not vary appreciably with $Q^2$. This enabled us to use our results for $a$ and $b$  presented in the previous section (see Fig.~\ref{fig:ab}). The proton spin structure function  $g_{1}^{p}$ was evaluated at the appropriate $x$ and $Q^2$ using the parametrization of \cite{GRSV}. Also note that while the values of $x$ and $Q^2$ are correlated for the HERMES data, the E154 Collaboration has evolved their data to the common scale $Q^2=5$ GeV$^2$. 
\begin{figure}[ht]
\begin{center}
\epsfig{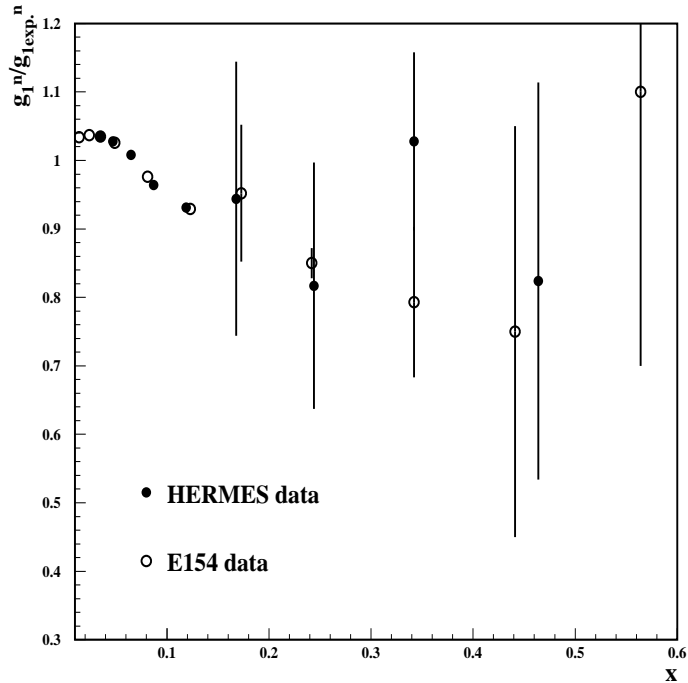}
\caption{The ratio 
$g_{1}^{n} /g_{1 \, {\rm exp}}^{n}$ based on Eq.~(\ref{ratio}), which  demonstrates how the HERMES \protect\cite{HERMES} and E154 \protect\cite{E154} values for $g_{1 \, {\rm exp}}^{n}$ should be corrected to include nuclear shadowing, antishadowing and the $\Delta$ isobar effects. The statistical uncertainty of 
$g_{1 \, {\rm exp}}^{n}$ contributes to the 
uncertainty of our predictions for $g_{1}^{n} /g_{1 \, {\rm exp}}^{n}$, which is shown by vertical lines.}
\label{fig:ratio}
\end{center}
\end{figure}

One can see from Fig.~\ref{fig:ratio} that in the region of nuclear shadowing, $10^{-4} \leq x \leq x_{0}$,  ignoring nuclear shadowing  would lead one to  overestimate $g_{1}^{n}$. For the lowest-$x$ experimental data points, this effect is of the order 4\%. At larger $x$, $x_{0} \leq x \leq 0.2$, the inclusion of nuclear antishadowing  increases  $g_{1}^{n}$. For instance, the increase is 7\% at $x \approx 0.12-0.13$, where the antishadowing correction is maximal. 
The influence of the $\Delta$ isobar on the extraction of $g_{1}^{n}$ from the $^3$He data is even larger: the experimental values for $g_{1}^{n}$ should be increased by as much as 15-25\%.

It is also interesting to note that 
the correction associated with the presence of the $\Delta$ isobar changes the value of Bjorken $x$, where $g_{1}^{n}$ changes sign. Indeed, as can be seen from Eq.~(\ref{ratio}), $g_{1}^{n}$ is larger than $g_{1 \, {\rm exp}}^{n}$ for $x > x_{0}$, i.e.  $g_{1}^{n}$ changes sign at smaller $x$ than $g_{1 \, {\rm exp}}^{n}$. In order to see the magnitude of this effect, 
we analyze Eq.~(\ref{ratio}) with $g_{1}^{p}$ and $g_{1}^{n}$ given by the parametrization of Ref.~\cite{GRSV}. Note that $g_{1}^{n}$ obtained in Ref.~\cite{GRSV} was fitted to the experimental data without the correction associated the $\Delta$ isobar and, thus, corresponds to $g_{1 \, {\rm exp}}^{n}$.
 Figure~\ref{fig:g1n} presents $g_{1}^{n}$ based on Eq.~(\ref{ratio}) as a solid curve and the free neutron spin structure function
$g_{1 \, {\rm exp}}^{n}$ as a dashed curve.
The two curves correspond to $Q^2$=4 GeV$^2$.
 \begin{figure}[ht]
\begin{center}
\epsfig{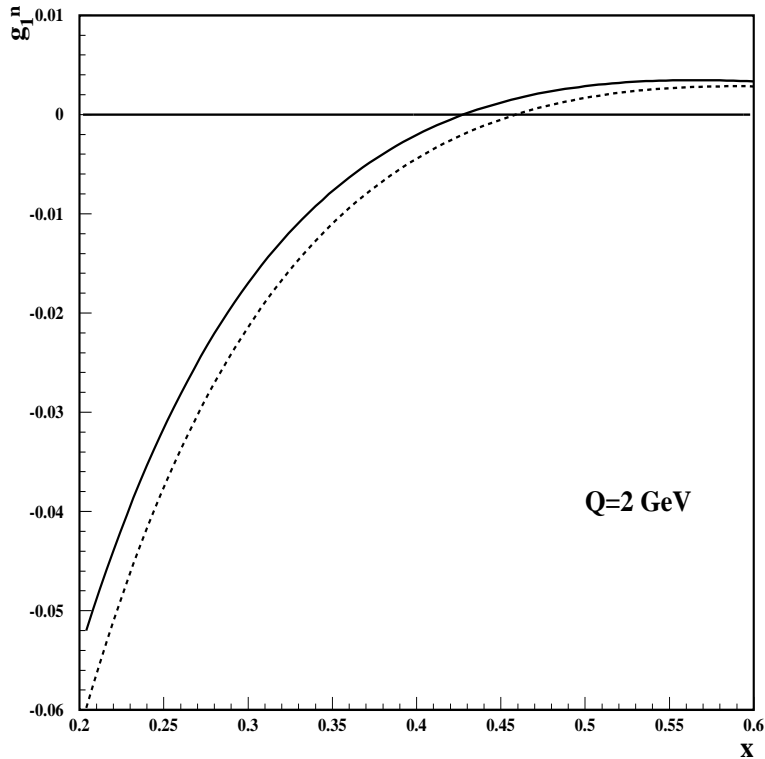}
\caption{The neutron spin structure function $g_{1}^{n}$ based on Eq.~(\ref{ratio}) (solid curve) compared to the case based on
 the parametrization of Ref.~\protect\cite{GRSV} (dashed curve).}
\label{fig:g1n}
\end{center}
\end{figure}
One can see from Fig.~\ref{fig:g1n} that for a given choice of $Q^2$ and shapes of $g_{1}^{n}$ and $g_{1}^{p}$,
 the presence of the $\Delta$ shifts the point where $g_{1}^{n}$ changes sign, from 0.46 to 0.43.

The effect of the $\Delta$ on the ratio $g_{1}^{n}/g_{1 \, {\rm exp}}^{n}$ is much more dramatic. If we form the ratio $g_{1}^{n}/g_{1 \, {\rm exp}}^{n}$ using the results presented in Fig.~\ref{fig:g1n}, its shape is quite similar to the tendency presented in Fig.~\ref{fig:ratio}: $g_{1}^{n}/g_{1 \, {\rm exp}}^{n}$ dips below unity for $0.2 \leq x < 0.4$ and rises above unity for $x > 0.5$.  However, the ratio $g_{1}^{n}/g_{1 \, {\rm exp}}^{n}$ exhibits extremely rapid changes from being large and negative to large and positive in the interval $0.4 < x < 0.5$, where $g_{1}^{n}$ changes sign. This effect is not seen in Fig.~\ref{fig:ratio}, where the discrete values of $g_{1\, {\rm exp}}^{n}$ are never close enough to zero. In the future, experimental studies of $g_{1\, {\rm exp}}^{n}$ near its zero would provide a very sensitive test of our model for the contribution of the $\Delta$ isobar to $g_{1}^{^3{\rm He}}$.

\section{$A_{1}^{n}$ from the $^3$He data at large $x$}

In this section we derive the expression necessary to extract the neutron asymmetry $A_{1}^{n}$ from the $^3$He data, which takes into account the presence of the $\Delta$ isobar in the $^3$He wave function. This calculation is motivated by the E99-117 experiment that is currently under way at TJNAF (USA) \cite{Meziani}. Using DIS on polarized $^3$He, the neutron asymmetry $A_{1}^{n}$ will be extracted from the $^3$He asymmetry $A_{1}^{^{\rm He}}$, which is measured with high accuracy in the large-$x$ region, $0.33 \leq x \leq 0.63$. 

The DIS asymmetry $A_{1}^{T}$ for any target $T$ is proportional to the spin structure function $g_{1}^{T}$:
\begin{equation}
g_{1}^{T}=\frac{F_{2}^{T}}{2 x(1+R)} A_{1}^{T} \ ,
\label{a1}
\end{equation}
where $R=(F_{2}^{T}-2x F_{1}^{T})/(2xF_{1}^{T})$ and  
$F_{1,2}^{T}$ are inclusive spin-averaged structure functions.   
It is assumed in Eq.~(\ref{a1}) that the transverse spin asymmetry, $A_{2}^{n}$, is negligibly small and that $R$ does not depend on the choice of target.

Applying this definition of $A_{1}^{T}$ to the $^3$He, proton and neutron targets and substituting into Eq.~(\ref{true}) where the terms proportional to $a$ and $b$ were omitted (we are interested in the large $x$ region, where shadowing and antishadowing are not present), one obtains for the neutron asymmetry $A_{1}^{n}$
\begin{equation}
A_{1}^{n}=\frac{F_{2}^{^3{\rm He}}}{P_{n} F_{2}^{n} (1+\frac{0.056}{P_{n}})}\Bigg(A_{1}^{^3{\rm He}}-2\frac{F_{2}^{p}}{F_{2}^{^3{\rm He}}}P_{p}A_{1}^{p} \bigg(1-\frac{0.014}{2 P_{p}}\bigg)\Bigg) \ .
\label{a1n}
\end{equation} 
Provided that the proton asymmetry, $A_{1}^{p}$, is constrained well by the experimental data, the largest theoretical uncertainty (which is of the order 10\%) in Eq.~(\ref{a1n}) comes from the uncertainty in the proton spin polarization $P_{p}$. Thus, we estimate that the uncertainty in the second term of  Eq.~(\ref{a1n}) and, thus, in the position of the point where $A_{1}^{n}$ has a zero, is of the order 10\%.

The terms proportional to $0.056$ and $0.014$ represent the correction to  $A_{1}^{n}$ associated with the $\Delta$ isobar. Both terms are important for the correct determination of  $A_{1}^{n}$. The term proportional to  $0.056$ decreases the absolute value of $A_{1}^{n}$ by about 6\%. Moreover, if $A_{1}^{^3{\rm He}}$ is negative, the second term proportional to $0.014$ would work in the same direction of decreasing of  $|A_{1}^{^3{\rm He}}|$. Since the term proportional to $0.014$ is always positive,  this means that  the true $A_{1}^{n}$ should turn positive at lower values of $x$ compared to the situation when the effect of the $\Delta$ is ignored (see Fig.~\ref{fig:g1n}).

\section{Summary and Conclusions}

We presented a comprehensive picture of nuclear effects relevant for DIS on polarized $^3$He, over a wide range of Bjorken $x$, $10^{-4} \leq x \leq 0.8$. These effects include nuclear shadowing and antishadowing, nucleon spin depolarization, Fermi motion and binding, the presence of the $\Delta$ isobar in the $^3$He wave function, and the off-shellness of the nucleons. For the first time, 
all the above effects were studied in a uniform fashion
using the ground-state wave function of $^3$He, which was obtained as a solution of the Faddeev equation with a separable version of the Paris nucleon-nucleon potential (PEST) with 5 channels. It is crucial to include all relevant nuclear effects for the proper determination of  the neutron spin structure function $g_{1}^{n}$ from the $^3$He data.  
In particular, we emphasized that the commonly used approximate expression for $g_{1}^{^3{\rm He}}$ based on Eq.~(\ref{conv2}), 
receives important corrections from  
the effects associated with  nuclear shadowing and antishadowing and  the $\Delta$ isobar (see Eq.~(\ref{true})). As as consequence, the values of  the neutron spin structure function, $g_{1}^{n}$, deduced from the $^3$He data by the E154 experiment at SLAC and the HERMES experiment at DESY should be corrected. Our results should be also taken into consideration in analysing the results of future DIS experiments on polarized $^3$He, such as, for instance, the E99-117 experiment at TJNAF. 
Our results are summarized below, starting from  the smallest $x$. 

At small values of Bjorken $x$, $10^{-4} \leq x \leq 0.2$, $g_{1}^{^3{\rm He}}$ is affected by nuclear shadowing and antishadowing as well as nucleon spin depolarization effects (see Eq.~(\ref{g1sh2})). As a result, the deviation from the approximate expression for  $g_{1}^{^3{\rm He}}$ given by Eq.~(\ref{conv2}) could be as large as 11-13\% at $x=10^{-3}$. This requires a 4\% decrease of the lowest-$x$  values for $g_{1}^{n}$ reported by the E154 and HERMES experiments. The effect of the antishadowing correction to $g_{1}^{^3{\rm He}}$ is somewhat smaller and works in the opposite direction: the experimental values for the extracted $g_{1}^{n}$ should be increased. For instance, the increase is 6\% at $x$=0.13. Note, however, that our treatment of  antishadowing is model-dependent and our predictions for the amount of antishadowing (and shadowing at $x$ close to $x_{0}$) depend crucially on the choice of $x_{0}$, the cross-over point between the nuclear shadowing and antishadowing regions. 
 
At larger $x$, $0.2 \leq x \leq 0.8$, the three principal nuclear effects are the nucleon spin  depolarization, the presence of the $\Delta(1232)$ resonance in the $^3$He wave function and Fermi motion and binding effects. The effect of 
the $\Delta$ works to decrease $g_{1}^{^3{\rm He}}$. For example, the decrease is of the order 12\% at $x=0.2$. The modification caused by the $\Delta$ is very significant at $x \approx 0.46$, where $g_{1}^{n}$ (in the particular parametrization of Ref.~\cite{GRSV}) is expected to change sign (for example, predictions for the shape of $g_{1}^{n}$ were derived in Refs.~\cite{MITbag} within the MIT bag model).
 In the region $0.2 \leq x \leq 0.8$, the E154 and HERMES values for $g_{1}^{n}$  should be increased by as much as 15-25\%.
Also, the effect associated with the $\Delta$ is expected to increase 
the neutron DIS asymmetry $A_{1}^{n}$, which will be measured by the E99-117 experiment at TJNAF.  As a result, the true $g_{1}^{n}$ should change sign at lower $x$.

The data files with the results presented in this work are available on request  from V. Guzey.

\section{Acknowledgements}
 
We would like to thank K. Tsushima for discussions concerning the results of Ref.~\cite{Steffens} and Z.-E. Meziani for raising the question of the extraction of $A_{1}^{n}$ from $A_{1}^{^3{\rm He}}$ related to the 99-117 experiment at TJNAF. This work was supported by the Australian Research Council, Adelaide University and U.S. Department of Energy.

\section{Appendix: Nuclear shadowing in polarized DIS on $^3$He}

In order to estimate nuclear shadowing in polarized DIS on $^3$He we use the standard Gribov-Glauber multiple scattering formalism (for a pedagogical 
review of the method, see \cite{Bauer}). The cross section for $h_{eff}$-$^3$He scattering with parallel helicities,
$\sigma_{h_{eff}A}^{\uparrow \uparrow}$ can be expressed through the nuclear profile function $\Gamma_{^3{\rm He}}^{\uparrow \uparrow}$:
\begin{equation}
\sigma_{h_{eff}A}^{\uparrow \uparrow}=2 Re \int d^2 b \Gamma_{^3{\rm He}}^{\uparrow \uparrow}(b) \ , 
\label{appen1}
\end{equation}
where $\vec{b}$ is a vector of the impact parameter, the distance between the projectile and the centre of the nucleus in the plane transverse to the direction of the incoming photon. The nuclear profile function $\Gamma_{^3{\rm He}}^{\uparrow \uparrow}$ is obtained as a series over nucleon spin-dependent profile functions $\Gamma_{i}(\vec{b}-\vec{r}_{i \perp})$ averaged with the ground-state wave function of $^3$He
\begin{eqnarray}
&&\Gamma_{^3{\rm He}}^{\uparrow \uparrow}=\langle \Psi_{^3{\rm He}}^{\uparrow}| \sum_{i}^{3} \sum_{s}\Gamma_{i}^{\uparrow s}(\vec{b}-\vec{r}_{i \perp}) \nonumber\\
&&-\sum_{i \neq j}^{3} \sum_{s_{1},s_{2}} \Gamma_{i}^{\uparrow s_{1}}(\vec{b}-\vec{r}_{i \perp})  \Gamma_{j}^{\uparrow s_{2}}(\vec{b}-\vec{r}_{j \perp}) \Theta (z_{j}-z_{i}) e^{i q_{\parallel} (z_{i}-z_{j})}  \nonumber\\
&&+\sum_{i \neq j \neq k}^{3}\sum_{s_{1},s_{2},s_{3}}  \Gamma_{i}^{\uparrow s_{1}}(\vec{b}-\vec{r}_{i \perp})  \Gamma_{j}^{\uparrow s_{2}}(\vec{b}-\vec{r}_{j \perp}) \Gamma_{k}^{\uparrow s_{3}}(\vec{b}-\vec{r}_{k \perp}) \nonumber\\
&&\times
\Theta (z_{j}-z_{i})
\Theta (z_{k}-z_{j}) e^{i q_{\parallel} (z_{i}-z_{k})}|\Psi_{^3{\rm He}}^{\uparrow}\rangle \ .
\label{a2}
\end{eqnarray} 
The helicity of the virtual photon is denoted by the first arrow in the superscripts; the helicity of the target nucleus is shown by an arrow next to the nuclear wave function.
 Since the helicities of the nucleons need not  be aligned with the helicity of the target, there are sums over helicities of the nucleons (symbolized by $s_{1}$, $s_{2}$ and $s_{3}$ in the superscripts). The subscript on the  $\Gamma$'s is designed to distinguish between the neutrons and protons.
 Positions of the nucleons with respect to the centre of the nucleus are given by transverse ($\vec{r}_{i \perp}$) and longitudinal ($z_{i}$) coordinates. The factors $e^{i q_{\parallel} (z_{i}-z_{j})}$ take into account the non-zero longitudinal momentum transferred to the nucleus, $q_{\parallel} \approx 2 m_{N} x$, where $m_{N}$ is the nucleon mass.

Using  time reversal one can show that the $\Theta$-functions in the double scattering terms of Eq.~(\ref{a2}) can be substituted by 1/2 and that the product of two $\Theta$-functions  in the triple scattering term can be substituted by 1/6.
 In addition, choosing the normalization of the $^3$He wave function that, for example, the first nucleon is the neutron (with coordinates $\vec{r}_{n}$) and the other two are protons (with coordinates $\vec{r}_{p}$ and $\vec{r}_{p^{\prime}}$), Eq.~(\ref{a2}) can be presented in the form
\begin{eqnarray}
&&\Gamma_{^3{\rm He}}^{\uparrow \uparrow}=\langle \Psi_{^3{\rm He}}^{\uparrow}| \sum_{s}\Big(\Gamma_{n}^{\uparrow s}(\vec{b}-\vec{r}_{n \perp})+2\Gamma_{p}^{\uparrow s}(\vec{b}-\vec{r}_{p \perp})\Big) \nonumber\\
&&-\sum_{s_{1},s_{2}}\Big(2 \Gamma_{n}^{\uparrow s_{1}}(\vec{b}-\vec{r}_{n \perp})  \Gamma_{p}^{\uparrow s_{2}}(\vec{b}-\vec{r}_{p \perp}) e^{i q_{\parallel} (z_{n}-z_{p})}+\Gamma_{p}^{\uparrow s_{1}}(\vec{b}-\vec{r}_{p \perp})  \Gamma_{p}^{\uparrow s_{2}}(\vec{b}-\vec{r}_{p^{\prime} \perp}) e^{i q_{\parallel} (z_{p}-z_{p^{\prime}})}\Big) \nonumber\\
&&+\sum_{s_{1},s_{2},s_{3}}  \Gamma_{n}^{\uparrow s_{1}}(\vec{b}-\vec{r}_{n \perp})  \Gamma_{p}^{\uparrow s_{2}}(\vec{b}-\vec{r}_{p \perp}) \Gamma_{p}^{\uparrow s_{3}}(\vec{b}-\vec{r}_{p^{\prime} \perp}) e^{i q_{\parallel} (z_{n}-z_{p^{\prime}})}|\Psi_{^3{\rm He}}^{\uparrow}\rangle \ .
\label{a3}
\end{eqnarray}

Each spin-dependent nucleon profile function is related to the spin-dependent $h_{eff}$-nucleon scattering cross section $\sigma_{N}^{\uparrow s}$ and the slope thereof, $B$:
\begin{equation}
\Gamma_{n,p}^{\uparrow s}(\vec{r}_{\perp})=\frac{\sigma_{n,p}^{\uparrow s}}{4 \pi B }e^{-\vec{r}_{\perp}^2 /(2B)} \ .
\label{a4}
\end{equation}

Combining Eqs.~(\ref{appen1},\ref{a3},\ref{a4}) one obtains for the $h_{eff}$-$^3$He spin-dependent scattering cross section
\begin{eqnarray}
&&\sigma_{h_{eff}A}^{\uparrow \uparrow}=\langle \Psi_{^3{\rm He}}^{\uparrow}|\sum_{s} \Big(\sigma_{n}^{\uparrow s}\hat{P}_{n}^{s}+2 \sigma_{p}^{\uparrow s} \hat{P}_{p}^{s} \Big) 
\nonumber\\
&&-\frac{1}{8 \pi B} \sum_{s_{1},s_{2}} \Big(2 \sigma_{n}^{\uparrow s_{1}} \sigma_{p}^{\uparrow s_{2}} \hat{P}_{n p}^{s_{1} s_{2}} +\sigma_{p}^{\uparrow s_{1}} \sigma_{p}^{\uparrow s_{2}} \hat{P}_{p p}^{s_{1} s_{2}} \Big) \nonumber\\
&&+\frac{1}{48 \pi^2 B^2} \sum_{s_{1},s_{2},s_{3}} \sigma_{n}^{\uparrow s_{1}} \sigma_{p}^{\uparrow s_{2}} \sigma_{p}^{\uparrow s_{3}} \hat{P}_{n p p}^{s_{1} s_{2} s_{3}}|\Psi_{^3{\rm He}}^{\uparrow} \rangle \ .
\label{a5}
\end{eqnarray}
Here the $\hat{P}$'s are projection operators onto one or several nucleons of $^3$He with particular helicities. The cross section for $h_{eff}$-$^3$He scattering with antiparallel helicities is obtained from Eq.~(\ref{a5}) by inverting the helicity of the target.

Next we introduce cross sections $\Delta \sigma$ and $\sigma$
\begin{eqnarray}
&&\sigma_{n,p}^{\uparrow \uparrow} \equiv \sigma+\frac{1}{2} \Delta \sigma_{n,p} \ , \nonumber\\
&&\sigma_{n,p}^{\uparrow \downarrow} \equiv \sigma-\frac{1}{2} \Delta \sigma_{n,p} \ .
\label{a6}
\end{eqnarray}
Here we do not distinguish between the spin-averaged cross sections for protons and neutrons.

Using Eqs.~(\ref{a5},\ref{a6}) the difference between the $h_{eff}$-$^3$He scattering cross sections with parallel and antiparallel helicities can be presented in the form
\begin{eqnarray}
&&\Delta \sigma_{h_{eff}A} \equiv \sigma_{h_{eff}A}^{\uparrow \uparrow}-\sigma_{h_{eff}A}^{\uparrow \downarrow}=P_{n}\Delta \sigma_{n}+2P_{p}\Delta \sigma_{p} \nonumber\\
&&-\frac{\sigma_{eff}}{4 \pi B}\Big(\Delta \sigma_{n} \Phi_{n}+\Delta \sigma_{p} \Phi_{p}\Big)+\frac{\sigma_{eff}^2}{48 \pi^2 (\alpha_{^3{\rm He}}+B)^2} \Delta \sigma_{n} \ .
\label{a7}
\end{eqnarray} 
Several remarks concerning Eq.~(\ref{a7}) are in order here. Firstly, $P_{n}$ and $P_{p}$ are effective proton and neutron spin polarizations defined by Eq.~(\ref{pnp}). Secondly, the nuclear shadowing correction to  $\Delta \sigma_{h_{eff}A}$, which is given by the second line of Eq.~(\ref{a7}), is determined by the effective spin-averaged cross section $\sigma_{eff}$ introduced in Sect.~\ref{sec:sha}. 
Thirdly, the nuclear shadowing correction due to  triple scattering, given by the last term in Eq.~(\ref{a7}), is small. As discussed in Sect.~\ref{sec:sha}, our numerical analysis demonstrated that the calculations with the exact, including higher partial waves, and highly simplified, where only the neutron is polarized, wave functions of $^3$He give very close results for the nuclear shadowing correction.  
Thus, to estimate the triple scattering contribution (last term in Eq.~(\ref{a7})),  it is safe to  use a simple Gaussian ansatz for the $^3$He ground-state wave function with $\alpha_{^3{\rm He}}=27$ GeV$^{-2}$
and assume that only the neutron is polarized \cite{FGS96}. Fourthly, the main effect of nuclear shadowing comes from the double scattering terms (proportional to $\Phi_{n}$ and $\Phi_{p}$) which need to be carefully evaluated.

 The functions $\Phi_{n}$ and $\Phi_{p}$ are defined as
\begin{eqnarray}
\Phi_{n}&=&\sum_{s_{1},s_{2}} \int \prod_{i} d^3 \vec{r}_{i} \Big(|\Psi_{\uparrow}(\vec{r}_{n},\uparrow; \vec{r}_{p}, s_{1}; \vec{r}_{p^{\prime}}, s_{2})|^2-|\Psi_{\uparrow}(\vec{r}_{n},\downarrow; \vec{r}_{p}, s_{1}; \vec{r}_{p^{\prime}}, s_{2})|^2\Big) \times \nonumber\\
&& e^{-(\vec{r}_{n \perp}- \vec{r}_{p \perp})^2/(4 B)} \cos \Delta (z_{n}-z_{p}) \ , \nonumber\\
\Phi_{p}&=&\sum_{s_{1},s_{2}} \int \prod_{i} d^3 \vec{r}_{i} \Big(|\Psi_{\uparrow}(\vec{r}_{n},s_{1}; \vec{r}_{p}, \uparrow; \vec{r}_{p^{\prime}}, s_{2})|^2-|\Psi_{\uparrow}(\vec{r}_{n},s_{1}; \vec{r}_{p}, \uparrow; \vec{r}_{p^{\prime}}, s_{2})|^2\Big) \times \nonumber\\
&& e^{-(\vec{r}_{n \perp}- \vec{r}_{p \perp})^2/(4 B)} \cos \Delta (z_{n}-z_{p}) + \nonumber\\
&&\sum_{s} \int \prod_{i} d^3 \vec{r}_{i} \Big(|\Psi_{\uparrow}(\vec{r}_{n},s; \vec{r}_{p}, \uparrow; \vec{r}_{p^{\prime}},\uparrow )|^2-|\Psi_{\uparrow}(\vec{r}_{n},s; \vec{r}_{p}, \downarrow; \vec{r}_{p^{\prime}}, \downarrow)|^2\Big) \times \nonumber\\
&& e^{-(\vec{r}_{p \perp}- \vec{r}_{p^{\prime} \perp})^2/(4 B)} \cos \Delta (z_{p}-z_{p^{\prime}}) \ .
\label{a8}
\end{eqnarray} 
Here $B=6$ GeV$^{-2}$ is the slope of the elementary $h_{eff}$-nucleon scattering cross section; $\Delta=q_{\parallel}=2 m_{N} x$. 
The used value for the slope $B$ requires discussion. It should be 
remembered that the elementary $h_{eff}$-nucleon scattering cross section is proportional to the diffractive electron-proton DIS cross section. Thus, $B$ is the slope of the diffractive electron-proton DIS cross section. The ZEUS collaboration measurement gives $B=7.2 \pm 1.1$ GeV$^{-2}$ \cite{ZEUS} in the HERA kinematics. Since  $B$ decreases slowly with decreasing energy, a slightly smaller value for $B$, $B=6$ GeV$^{-2}$, seems to be more appropriate for the kinematics of fixed target experiments on polarized DIS on nuclear targets.

 For the ground-state wave function of $^3$He we used the one obtained by solving the Faddeev equations with the PEST two-nucleon interaction potential including 5 channels \cite{Bissey01}. 

Using the relation between the spin structure function $g_{1}^{^3{\rm He}}$ and the difference of the cross sections, $\Delta \sigma_{h_{eff}A}$ (see Eq.~(\ref{g1})),
 one can find the most complete expression for the $^3$He spin structure function $g_{1}^{^3{\rm He}}$ (see Eqs.~({\ref{g1sh},\ref{g1sh2}))
\begin{eqnarray}
&&g_{1}^{^3{\rm He}}=\int_{x}^{A} \frac{dy}{y} \Delta f_{n/^3{\rm He}}(y)\tilde{g}_{1}^{n}(x/y)+\int_{x}^{A} \frac{dy}{y} \Delta f_{p/^3{\rm He}}(y)\tilde{g}_{1}^{p}(x/y) \nonumber\\
&&-0.014\Big(\tilde{g}_{1}^{p}(x)-4\tilde{g}_{1}^{n}(x)\Big)-a^{sh}(x) g_{1}^{n}(x)-b^{sh}(x) g_{1}^{p}(x)  \ ,
\label{a9}
\end{eqnarray}
where 
\begin{eqnarray}
&&a^{sh}(x,Q^2)=\frac{\sigma_{eff}}{4 \pi B} \Phi_{n}-\frac{\sigma_{eff}^2}{48 \pi^2 (\alpha_{^3{\rm He}}+B)^2} \, \nonumber\\
&&b^{sh}(x,Q^2)=\frac{\sigma_{eff}}{4 \pi B} \Phi_{p} \ .
\label{ab}
\end{eqnarray}
In Eq.~(\ref{a9}), we replaced the single scattering terms proportional to $P_{n}$ and $P_{n}$ by their generalization in terms of the convolution with the off-shell nucleon structure functions. Also, the effects associated with the presence of the $\Delta$ isobar were included.

\end{document}